\documentclass[conference]{IEEEtran}
\IEEEoverridecommandlockouts
\usepackage[font=footnotesize,labelfont=bf]{subcaption}
\usepackage{cite}
\usepackage{amsmath,amssymb,amsfonts}
\usepackage{algorithmic}
\usepackage{graphicx}
\usepackage{textcomp}
\usepackage{xcolor}
\usepackage[autostyle=true]{csquotes} 
\usepackage[english]{babel}
\usepackage{hyperref}
\usepackage{multirow} 
\usepackage{makecell}
\usepackage{float}
\def\BibTeX{{\rm B\kern-.05em{\sc i\kern-.025em b}\kern-.08emT\kern-.1667em\lower.7ex\hbox{E}\kern-.125emX}}
    
\begin{document}
\title{IntelliCardiac: An Intelligent Platform for Cardiac Image Segmentation and Classification}
\author{
    \IEEEauthorblockN{
        Ting Yu Tsai\textsuperscript{1},
        An Yu\textsuperscript{1},   
        Meghana Spurthi Maadugundu\textsuperscript{1},
        Ishrat Jahan Mohima\textsuperscript{1}, \\
        Umme Habiba Barsha\textsuperscript{1}, 
        Mei-Hwa F. Chen\textsuperscript{1},
        Balakrishnan Prabhakaran\textsuperscript{1},
        Ming-Ching Chang\textsuperscript{1,*} \thanks{* Corresponding author.}
    }
    \IEEEauthorblockA{
        \textsuperscript{1}University at Albany, State University of New York,\\
        {\tt \small \{ttsai2,ayu,mmaadugundu,imohima,hbarsha,mchen,bprabhakaran,mchang2\}@albany.edu}\\
    }
}

\maketitle
\begin{abstract}
Precise and effective processing of cardiac imaging data is critical for the identification and management of the cardiovascular diseases. We introduce IntelliCardiac, a comprehensive, web-based medical image processing platform for the automatic segmentation of 4D cardiac images and disease classification, utilizing an AI model trained on the publicly accessible ACDC dataset. The system, intended for patients, cardiologists, and healthcare professionals, offers an intuitive interface and uses deep learning models to identify essential heart structures and categorize cardiac diseases. The system supports analysis of both the right and left ventricles as well as myocardium, and then classifies patient's cardiac images into five diagnostic categories: dilated cardiomyopathy, myocardial infarction, hypertrophic cardiomyopathy, right ventricular abnormality, and no disease.
IntelliCardiac combines a deep learning-based segmentation model with a two-step classification pipeline. The segmentation module gains an overall accuracy of 92.6\%. The classification module, trained on characteristics taken
from segmented heart structures, achieves 98\% accuracy in five categories. These results exceed the performance of the existing state-of-the-art methods that integrate both segmentation and classification models. IntelliCardiac, which supports real-time visualization, workflow integration, and AI-assisted diagnostics, has great potential as a scalable, accurate tool for clinical decision assistance in cardiac imaging and diagnosis. The developed app can be
accessed at \url{https://github.com/tiffany9056/IntelliCardiac}.
\end{abstract}


\section{Introduction}

Cardiovascular diseases (CVDs) remain the leading cause of mortality worldwide~\cite{WHO2023}, highlighting the need for accurate, efficient diagnostic tools. Cardiac magnetic resonance imaging (MRI) is a key modality for assessing cardiac function and structure, providing rich, time-resolved 4D data~\cite{Isensee2021nnUNet}. However, manual analysis is time-consuming, variable, and ill-suited to fast-paced clinical settings~\cite{Karamitsos2009}.

While deep learning has advanced cardiac image analysis—particularly in segmentation and disease classification~\cite{Pan2024S2S2}—most existing methods rely on 2D or 3D MRI data, often overlooking important temporal dynamics between frames. Available web platforms~\cite{ziegler2020open} typically focus on image visualization alone, lacking real-time, user-accessible tools that integrate processing, visualization, and diagnostic support in a single workflow.

\begin{figure}[t]
\centerline{\includegraphics[width=0.5\textwidth]{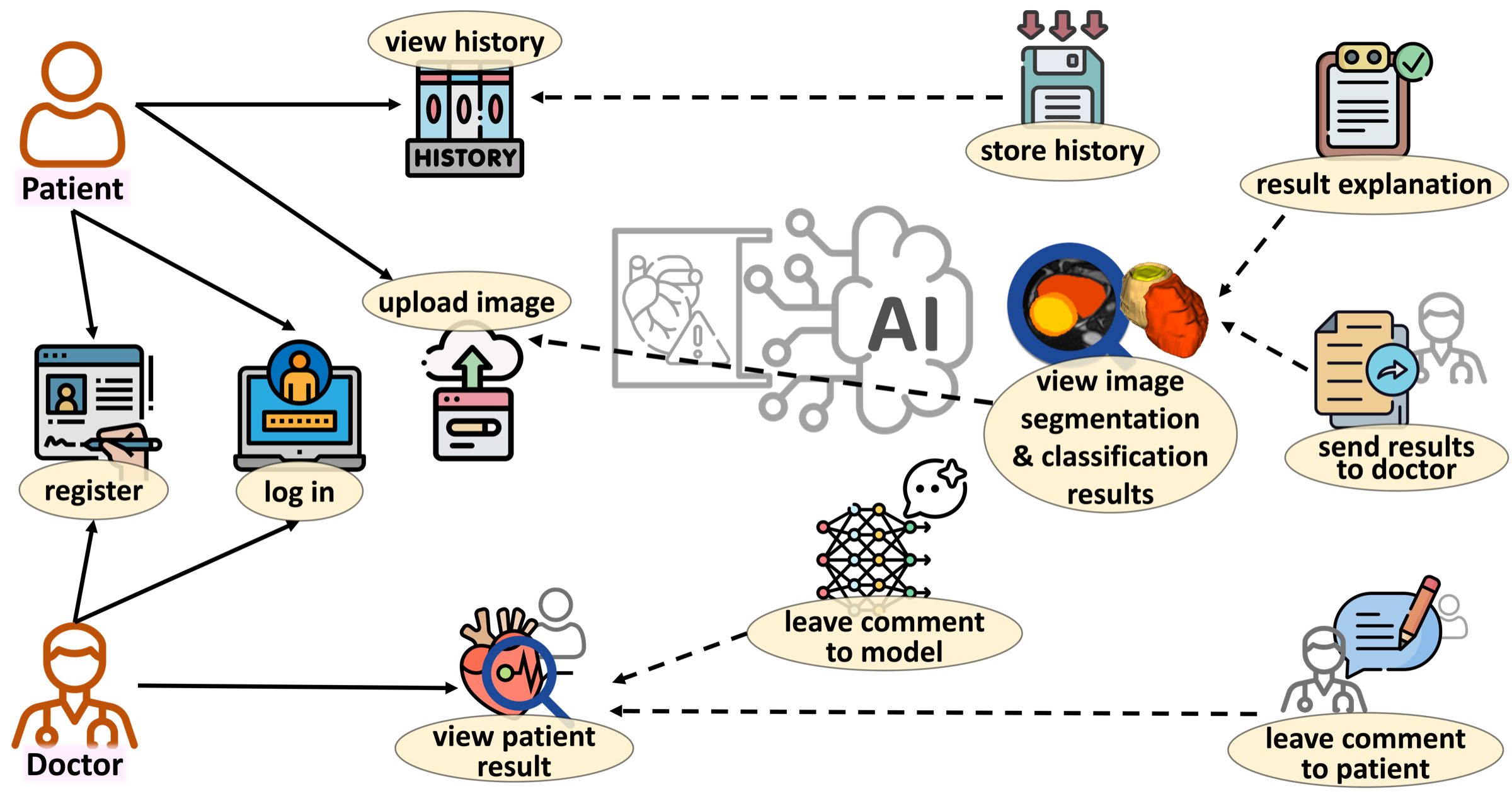}}
\vspace{-1mm}
\caption{Overview of IntelliCardiac system pipeline.}
\vspace{-1mm}
\label{fig:OverviewIntelliCardiac}
\end{figure}

To tackle the challenges, we present IntelliCardiac, a novel, web-based cardiac image processing platform that integrates deep learning models for automatic 4D image segmentation and disease classification. The system is designed with usability and scalability in mind, offering a seamless interface for healthcare professionals, researchers, and patients. IntelliCardiac incorporates:
(i) a deep learning-based segmentation model targeting both the right and left ventricles as well as myocardium,
(ii) a two-stage classification pipeline to assign patient's cardiac images into five diagnostic categories, and
(iii) a real-time web interface that enables visualization, interaction, and integration into clinical workflows. An overview of the IntelliCardiac system pipeline is shown in Figure~\ref{fig:OverviewIntelliCardiac}.

Models were trained and evaluated on the publicly available ACDC dataset~\cite{Bernard2018}, IntelliCardiac achieves 92.6\% segmentation accuracy and 98\% classification accuracy—outperforming prior state-of-the-art methods~\cite{Wolterink2018, Wibowo2022, Zheng2019, Khened2018} that combine both tasks. In addition to its technical performance, IntelliCardiac represents a significant step toward practical AI deployment in cardiac diagnostics by bridging the gap between advanced machine learning and real-world clinical utility.

The contributions of this paper are:
\begin{itemize}
    \item \textbf{An end-to-end web-based platform}: We present IntelliCardiac, a user-friendly system that integrates 4D cardiac MRI segmentation and classification for real-time interaction and clinical support.
    \item \textbf{A joint deep learning pipeline}: Our method combines high-accuracy segmentation of cardiac structures with a two-stage classification model, achieving 92.6\% and 98\% accuracy respectively on the ACDC dataset, outperforming existing approaches.
    \item \textbf{Clinical readiness and extensibility}: IntelliCardiac bridges the hole between AI research and real-world deployment, offering a scalable framework for cardiac diagnosis that can be extended to other modalities and disease categories.
\end{itemize}

\section{Related Work}
\noindent\textbf{Cardiac Image Analysis} Deep learning has advanced cardiac MRI analysis, particularly in segmentation and disease classification. U-Net and its variants~\cite{ronneberger2015u} for segmenting the myocardium and ventricles, with benchmarks like ACDC~\cite{Bernard2018} driving progress. Recent models enhance performance on 4D sequences by incorporating attention~\cite{oktay2018attention} and spatiotemporal features~\cite{shen2024spatiotemporal}. For disease classification, many approaches extract features from segmented structures and apply classifiers such as random forests~\cite{Breiman2001}, SVMs~\cite{Hearst1998}, or deep networks~\cite{pedersen2007circular}.End-to-end models~\cite{duda1972use} that combine segmentation and classification have emerged but often lack modularity and real-time capability. Most pipelines remain task-specific and are not designed for integrated, clinician-friendly applications~\cite{kikinis20133d}.

\noindent \textbf{Web-based Medical Imaging Platforms} Interest is growing in web-based AI tools for remote diagnostics and telemedicine. Platforms like OHIF Viewer~\cite{ziegler2020open} and 3D Slicer~\cite{kikinis20133d} offer basic visualization but lack deep integration with advanced AI pipelines. A few recent studies have proposed task-specific platforms for AI-assisted diagnosis~\cite{chatterjee2025cloud}, but these are rarely open, end-to-end, and designed with real-time processing in mind. Our proposed system, IntelliCardiac, addresses these limitations by offering a unified, web-based platform that integrates deep learning's state-of-the-art models for the segmentation and classification, supports real-time interaction, and is designed for practical use in clinical settings helping in early and efficient disease detection.

\begin{figure*}[t]
    \centerline{\includegraphics[width=1\textwidth]{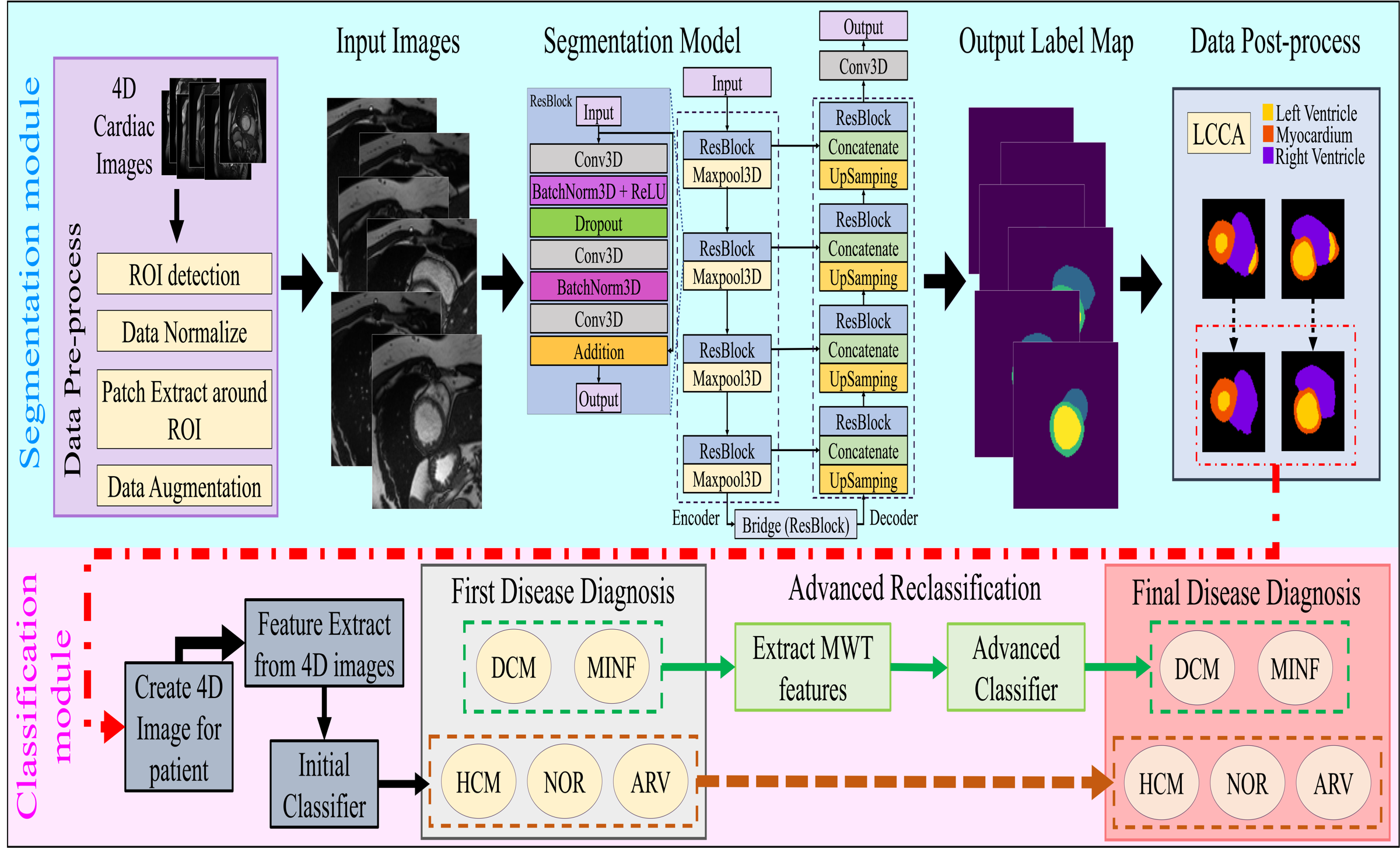}}
    \caption{Proposed AI model architecture consisting of segmentation (top) and two-stage classification (bottom).}
    \vspace{-4mm}
    \label{fig:proposed_architecture}
\end{figure*}

\section{Methodology}
\subsection{Web platform design}
Our system is designed around two primary user roles: Patients and Doctors. The core feature is a secure web service where patients upload 4D cardiac MRI images. Upon submission, the system automatically analyzes the MRI, segments key heart structures, and classifies the heart condition into one of the five disease categories using the AI model. For the predicted classification label, an interpretative explanation of the result is provided to the patient. This framework serves as a preliminary screening tool to assist professionals in efficiently managing higher patient volumes.

The user interface is designed for simplicity and ease of use, allowing most users to register and navigate the platform without prior training. Patients follow a straightforward workflow: they submit MRI images via the ``Upload Image" page, view diagnostic results and insights on the ``Test Evaluation" page, and can review all previous tests through the ``Check History" page, with options to delete results or adjust privacy settings. Patients may grant doctors permission to access specific reports and receive real-time notifications when comments are added. The ``Profile" page allows patients to manage personal information. Doctors can access all shared patient records through the ``View All Tests" page, comment on individual cases via the ``View Patient Result" page, and update their own profiles. The ``Our Doctors" page lists all registered doctors with links to their external websites, providing additional background. All user actions are protected by session-based authentication and encrypted credentials, ensuring robust security and data privacy.

\subsection{Segmentation module}

\subsubsection{Data Pre-processing}
To ensure robust learning and enhance the generalizability of our segmentation model, we used a complete preprocessing pipeline that included spatial normalization, Region of interest (ROI) extraction, data augmentation, and structured cropping. Our ROI detection strategy plays a critical role in maximizing segmentation accuracy and maintaining consistency across patients with variable anatomy and scan alignment. To localize the cardiac region, we compute the temporal standard deviation of pixel intensities at each spatial location to highlight dynamic cardiac structures. Circular contours in axial slices are then detected using the Canny edge detector~\cite{canny1986computational} and Circular Hough Transform~\cite{duda1972use, pedersen2007circular}, assuming the left ventricle (LV) has a roughly circular cross-section. A Gaussian voting method generates a likelihood surface, and its peak is selected as the LV center. Around this center, we extract a fixed-size $128 \times 128$ patch, combined with our logical-depth cropping strategy, which generates multiple temporally aware crops when the depth exceeds the target size. This preprocessing framework allows the model to capture the full spatial-temporal dynamics of the cardiac cycle.

For better generalization, we used a sequence of data augmentations on the images and labels during training. These include random flips, in-plane rotations, spatial translations, and affine shearing. This pipeline ensures that the segmentation model's input is uniform, anatomically centered, and diverse enough to capture the observed inter-patient heterogeneity.

\subsubsection{Network Architecture for Segmentation}
As illustrated in the segmentation module section of Figure~\ref{fig:proposed_architecture}, we implement a 3D U-Net-style encoder-decoder architecture along with the residual connections for cardiac segmentation. The network is constructed using stacked ResBlock modules, each of which contains two convolutional layers, batch normalization, ReLU activation, dropout, and a residual skip connection.

The encoder is composed of four ResBlock layers, among them every layer is followed by max pooling to downsample spatially. The encoder and decoder are joined by a deeper connection of ResBlock. Inside the decoder, skip-connected features from the encoder are linked with the unsampled feature maps that have been upsampled using trilinear interpolation. The segmentation is improved by passing each concatenated feature map through a ResBlock. The prediction is outputted with four channels in the final convolution, which corresponds to the segmentation's final label map.

\subsubsection{Loss Function}
We adopt the Focal Dice Loss~\cite{Wang2018} in multi-class 3D segmentation, which extends the standard Dice loss by applying an exponent $\beta$ to focus training on difficult predictions. The class-wise formulation is given by:
\[
\mathcal{L}_{\text{FocalDice}}^{(c)} = 1 - \left( \frac{2 \cdot \sum \mathbf{P}_c \cdot \mathbf{G}_c + \epsilon}{\sum \mathbf{P}_c^2 + \sum \mathbf{G}_c^2 + \epsilon} \right)^{\frac{1}{\beta}}
\]
where $\mathbf{P}_c$ and $\mathbf{G}_c$ are the predicted probabilities and one-hot encoded ground truth for class $c$, respectively, and $\epsilon$ acts as a small constant for numerical stability.

The overall loss is computed as a weighted average over all $C$ classes:
\[
\mathcal{L}_{\text{total}} = \frac{1}{\sum_{c=1}^{C} w_c} \sum_{c=1}^{C} w_c \cdot \mathcal{L}_{\text{FocalDice}}^{(c)}
\]

We have a contribution lies in the dynamically weighted Focal Dice Loss. Instead of using fixed weights, we update the class weights at each epoch based on the inverse of the average Dice scores from the previous epoch:
\[
w_c \propto 1 - \text{Dice}_c
\]
This inspires the model to focus on underperforming classes while training, effectively addressing class imbalance and stabilizing performance across all regions of interest.

\begin{figure*}[t]
    \centerline{\includegraphics[width=1.0\textwidth]{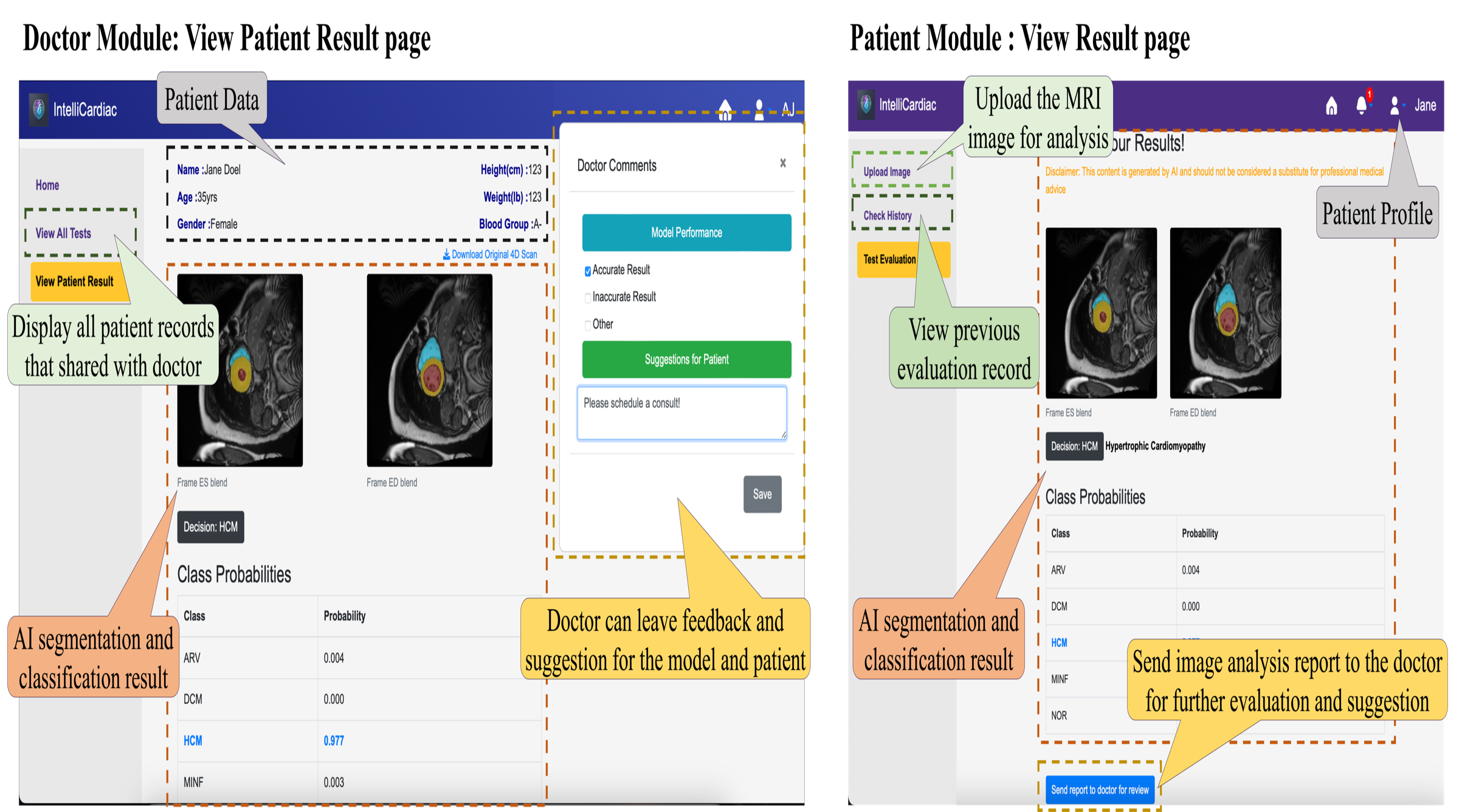}}
    \caption{User interface overview of IntelliCardiac: the left shows the doctor’s view with patient results and diagnostic insights; the right shows the patient’s view with AI-generated cardiac evaluation results.\vspace{1em}}
    \vspace{-5mm}
    \label{fig:platform_pages}
\end{figure*}

\begin{table}[t]
\centering
\caption{Summary of the features extracted from the segmentation masks at end-diastole (ED) and end-systole (ES) for cardiac disease classification. These features capture ventricular volumes, myocardial mass, ejection fractions, volume ratios, and myocardial wall thickness (MWT) statistics.}
\label{tab:classification_features}
\begin{tabular}{p{3.9cm}|p{4.1cm}}
    \hline
    \multicolumn{2}{c}{\textbf{Feature Descriptions}}\\
    \hline
    Left ventricle volume at ED \& ES & Right ventricle volume at ED \& ES \\
    Myocardial volume at ED \& ES     & Ejection fraction at LV \& RV \\
    $LV/RV$ volume ratio at ED \& ES    & $Myo/LV$ volume ratio at ED \& ES \\
    Max of mean MWT at ED \& ES       & Std of std MWT at ED \& ES \\
    Mean of std MWT at ED \& ES       & Std of mean MWT at ED \& ES \\ \hline
\end{tabular}
\vspace{-7pt}
\end{table}
\subsubsection{Data Post-processing}
Following obtaining of the raw segmentation predictions, post-processing is implemented to improve anatomical consistency and return the outputs to their original spatial context. Initially, we conduct Largest Connected Component Analysis (LCCA)~\cite{Samet1988, Dillencourt1992} independently on each predicted class (excluding background). We effectively eliminate small, isolated false positives by identifying all connected components in the 3D volume for each class and retaining only the most massive one. The structural coherence of the segmented regions is enhanced by this step. Subsequently, we revert the segmentation from the fixed-size cropped region to the original image dimensions. Lastly, the ED and ES segmentations that have been processed are stacked along the temporal axis to create a 4D volume. This volume is used as input for the downstream classification model.

\subsection{Classification module}
The target of the automated cardiac or heart disease detection task is to classify each cine MRI scan to one of five diagnostic categories: dilated cardiomyopathy (DCM), myocardial infarction (MINF), hypertrophic cardiomyopathy (HCM), normal (NOR) or abnormal right ventricle (ARV).

\subsubsection{Feature Extraction}
In order to perform this classification, we extracted a comprehensive set of 20 features from the 4D segmentation maps at end-diastole (ED) as well as end-systole (ES), inspired by the feature-based approach suggested by Isensee {\em et al.}~\cite{Isensee2018}. All the features are listed in Table~\ref{tab:classification_features}. These features are indicative of the heart's volumetric, functional, and morphological characteristics. First, we calculated the volume of the left and right ventricle along with myocardium (Myo) in both the ED and ES phases. To find this we multiplied the number of voxels assigned to each class by the voxel size. Second, we performed an analysis of the myocardial wall thickness by measuring the Euclidean distance between the outer and inner contours of the myocardium in each short-axis slice. We generated statistical summaries of the wall thickness distributions for both ED and ES. The summaries included the maximum and standard deviation of the slice-wise means, also the mean and standard deviation of the slice-wise standard deviations.

\subsubsection{Network Architecture for Classification}
In order to classify cardiac conditions using the extracted features, we developed a two-stage classification framework that integrates Random Forest~\cite{Breiman2001} and Support Vector Machine (SVM)~\cite{Hearst1998} classifier.

\textbf{Initial Classifier:}  
During the first stage, a Random Forest classifier was trained by us with the complete set of extracted cardiac features and all training cases. The model generates one of five diagnostic categories.

\textbf{Advanced Classifier:}  
Clinically, DCM and MINF show similar volumetric patterns but vary in myocardial wall shape. Typically exceeding in the end-systolic (ES) phase, myocardial wall thickness (MWT) is thinnest in end-diastole (ED)~\cite{Karamitsos2009}. In healthy individuals, MWT segments vary consistently. Localized infarct zones cause MINF wall thickness patterns to be highly uneven. DCM segments have uniformly thin walls. The fact that general-purpose classifiers struggle to capture these nuanced distinctions supports the need for a focused expert model. We constructed an advanced classifier targeted to improving MINF/DCM predictions to resolve these ambiguities. Support Vector Machine with RBF kernel, trained on two clinically significant MWT features: Samples predicted as MINF or DCM by the initial classifier were re-evaluated using the maximum of slice-wise mean wall thickness and the mean of slice-wise standard deviations measured during the ES phase.

\textbf{Final Classification Decision:}  
The revised predictions replaced initial MINF/DCM predictions. The final classification result was formed by incorporating the corrected predictions from advanced classifier into the full prediction array.

\section{Application and Experiment}

\subsection{Application}
The tech stacks used to develop this application are Express.js, Node.js, EJS, Python, and MongoDB. This architecture enables scalable, efficient processing and secure interaction between AI modules and users. Figure~\ref{fig:platform_pages} presents a two-page visual overview of the platform interface, highlighting key functions and features available to both user groups.

\subsubsection{Doctor Module}
This module supports cardiologists in reviewing and responding to patient reports. Its key functionalities are as follows:

\begin{itemize}
    \item \textbf{Authentication:} Secure registration and login for doctors.
    \item \textbf{Patient Report Access:} Patient reports include their demographic data, AI-generated segmented cardiac images, predicted classification labels and their associated probabilities.
    \item \textbf{Recommendations to Patients:} Doctors can submit and edit suggestions for follow-up care based on the reports.
    \item \textbf{Model Feedback:} Doctors may provide feedback on the AI's output to aid future model improvement.
\end{itemize}

\subsubsection{Patient Module}
The patient module enables users to manage their health data, interact with the AI diagnostic tool and optionally receive clinical suggestions from cardiologists. Core functionalities include:

\begin{itemize}
    \item \textbf{Authentication:} Patients can register and log in to manage their profiles.
    \item \textbf{MRI Upload and Evaluation:} Cardiac MRI scans are uploaded and analyzed by the AI model to generate diagnostic reports containing segmented cardiac images, predicted classification labels and associated probabilities.
    \item \textbf{Result Explanation:} Detailed explanation of the predicted classification label can be viewed by the patient to better understand the generated report.
    \item \textbf{Report Submission:} Patients may selectively share reports with doctors. Access is controlled by user-defined privacy settings.
    \item \textbf{Access to Recommendations:} Patients receive notifications when doctors provide feedback.
    \item \textbf{Record History and Deletion:} Users can view previous evaluations and delete their records from the platform.
    
\end{itemize}


\begin{figure}[t]
\centerline{\includegraphics[width=0.5\textwidth]{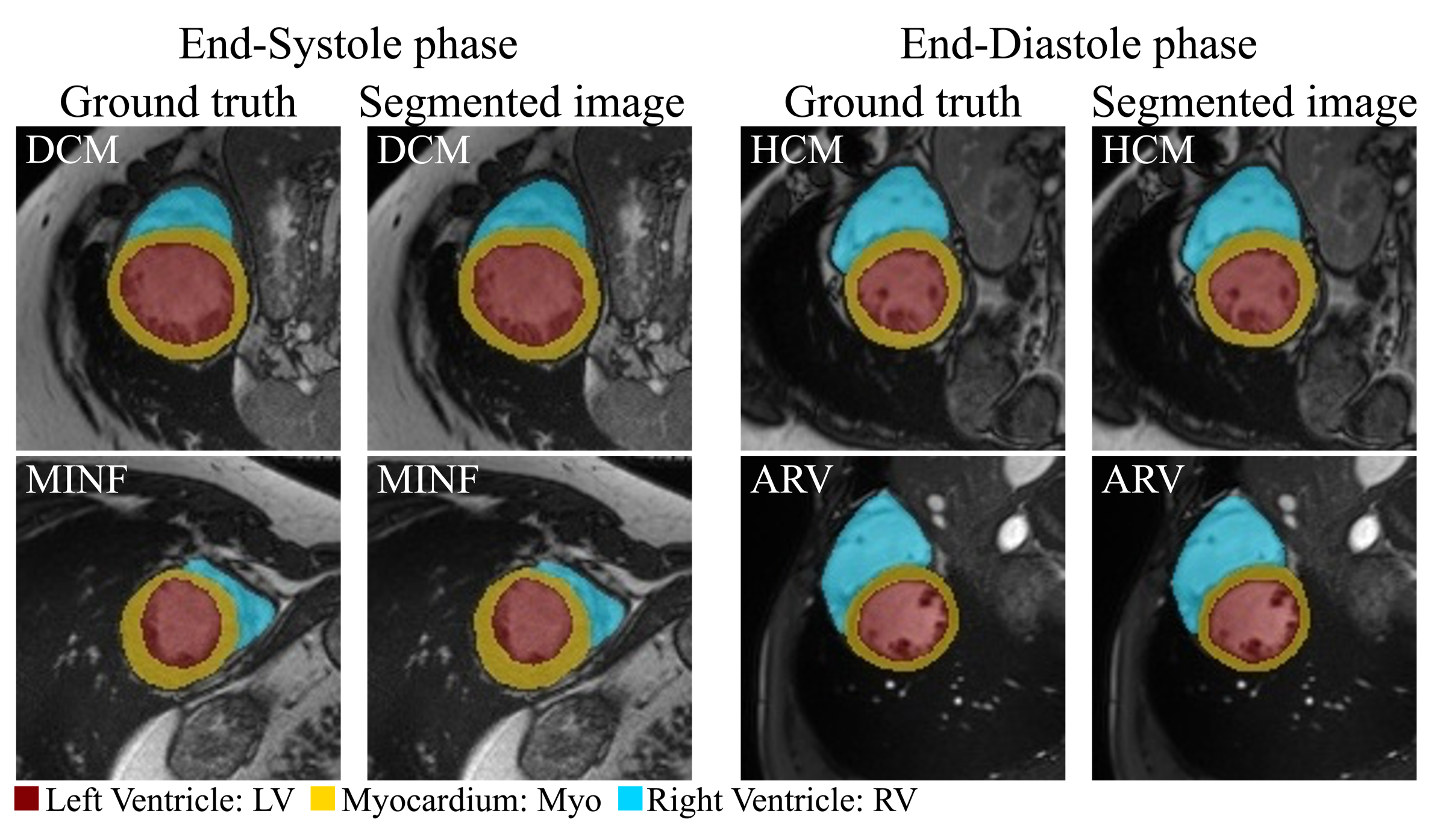}}
\vspace{-2mm}
\caption{Segmentation and classification results from proposed model on sample images from the ACDC dataset. The ground truth is displayed in the first and third columns, while the corresponding segmented images and classification results are presented in the second and fourth columns.}
\vspace{-2mm}
\label{fig:demo_output}
\end{figure}

\begin{figure}[t]
\centerline{\includegraphics[width=0.5\textwidth]{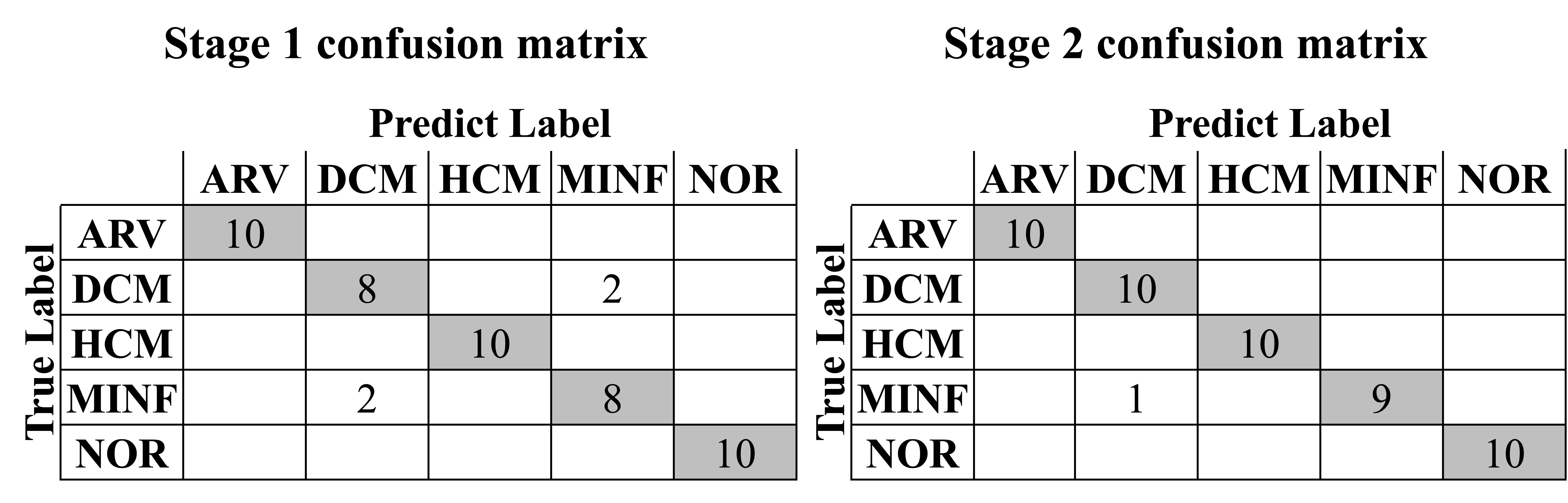}}
\vspace{-1mm}
\caption{Confusion matrices before (right) and after (left) expert refinement in IntelliCardiac classification module.}
\vspace{-2mm}
\label{fig:confusion_matrices}
\end{figure}

\begin{table}[t]
\caption{Model Segmentation Performance Comparison on the ACDC Dataset. The evaluation metrics are mean Dice scores (\%) over individual heart substructures include RV (right ventricle), LV (left ventricle) and Myo (myocardium). The bolded outcomes shown here are the best ones.}
\label{tab:Results_segmentation}
\centerline{
    \scalebox{0.9}{
    \begin{tabular}{c|c|c|c|c}
        \hline
        \textbf{Methods}                    & \textbf{Avg. DSC} $\uparrow$ & \textbf{RV} $\uparrow$  & \textbf{Myo} $\uparrow$ & \textbf{LV} $\uparrow$  \\ \hline
        TransUNet \cite{Chen2021TransUNet} & 89.71                        & 88.86                   & 84.53                   & 95.73                   \\  
        Swin-Unet \cite{Cao2021SwinUnet}   & 90.00                        & 88.55                   & 85.62                   & 95.83                   \\ 
        S2S2      \cite{Pan2024S2S2}       & 90.40                        & 88.95                   & 86.16                   & 96.07                   \\ 
        SegFormer3D \cite{Perera2024}      & 90.96                        & 88.50                   & 88.86                   & 95.53                   \\ 
        nnUNet    \cite{Isensee2021nnUNet} & 91.61                        & 90.24                   & 89.24                   & 95.36                   \\ 
        nnFormer  \cite{Zhou2021nnFormer}  & 92.06                        & 90.94                   & 89.58                   & 95.65                   \\ 
        EMCAD     \cite{Rahman2024EMCAD}   & 92.12                        & 90.95                   & 89.68                   & \textbf{96.26}          \\ \hline
        \textbf{Our IntelliCardiac}        & \textbf{92.56} (+0.44)       & \textbf{92.27} (+1.32)  & \textbf{90.33} (+0.65)  & 95.09                   \\ \hline
    \end{tabular}
    }
}
\vspace{-1mm}
\end{table}

\begin{table}[t]
\caption{Model performance comparison on the ACDC dataset for segmentation and classification. The evaluation metrics for segmentation include mean Dice scores (\%) averaged over heart substructures. Classification performance is reported as overall accuracy (\%).}
\label{tab:Results_seg_clas}
\centerline{
    \scalebox{0.7}{
    \begin{tabular}{c|c|c|c|c|c}
        \hline
                                                    & \multicolumn{4}{c|}{\textbf{Segmentation}}                                                     & \textbf{Classification}     \\ \hline
        \textbf{Methods}                             & \textbf{Avg. DSC} $\uparrow$ & \textbf{RV} $\uparrow$  & \textbf{Myo} $\uparrow$ & \textbf{LV} $\uparrow$ & \textbf{Accuracy} $\uparrow$\\ \hline
        Wolterink {\em et al.} \cite{Wolterink2018} & 89.30                        & 88                      & 87                      & 93                     & 86\%                          \\  
        Wibowo {\em et al.}    \cite{Wibowo2022}    & 90.89                        & 89.81                   & 89.69                   & 93.08                  & 92\%                          \\ 
        Zheng {\em et al.}     \cite{Zheng2019}     & 88.33                        & 87                      & 84                      & 94                     & 94\%                          \\ 
        Khened {\em et al.}    \cite{Khened2018}    & 88.33                        & 87                      & 86                      & 92                     & 90\%                          \\ \hline
        \textbf{Our IntelliCardiac}                 & \textbf{92.56} (+1.67)       & \textbf{92.27} (+2.46)  & \textbf{90.33} (+0.64)  & \textbf{95.09} (+1.09) & \textbf{98\%} (+4\%)          \\ \hline
    \end{tabular}
    }
}

\vspace{-3mm}
\end{table}

\begin{table}[t]
\caption{Ablation study on various configurations of the segmentation backbone. The evaluation metric for segmentation is the mean Dice score (\%), averaged over individual heart substructures: RV (right ventricle), LV (left ventricle) and Myo (myocardium). The proposed dynamically weighted Focal Dice Loss with ROI detection is used.}
\label{tab:Results_ablation_study}
\centerline{
    \scalebox{1.0}{
    \begin{tabular}{c|c|c|c|c|c}
        \hline
        \multirow{2}{*}{\makecell{\textbf{Loss function /}\\\textbf{ROI detection}}}
                               & \multicolumn{4}{c|}{\textbf{Segmentation}} & \textbf{Classification} \\ \cline{2-6}
                               & \textbf{Avg.}  & \textbf{RV}    & \textbf{Myo}   & \textbf{LV}    & \textbf{Initial / Final} \\ \hline
        Cross Entropy          & 92.21          & 92.29          & 89.78          & 94.56          & 90\% / 96\%                              \\ 
        Dice                   & 92.45          & 92.25          & 90.22          & 94.88          & 92\% / 96\%                              \\ 
        Focal Dice             & 92.45          & 92.24          & 90.20          & 94.92          & 88\% / 92\%                              \\ 
        Ours (no ROI)          & 92.17          & 92.17          & 90.18          & 94.15          & 90\% / 94\%                              \\ \hline
        \textbf{Ours (w/ ROI)} & \textbf{92.56} & \textbf{92.27} & \textbf{90.33} & \textbf{95.09} & 92\% / \textbf{98\%}                     \\ \hline
    \end{tabular}
    }
}
\vspace{-3pt}
\end{table}

\subsection{Segmentation and Classification Model Experiments and Results}
The ACDC dataset was utilized to train our 3D residual U-Net segmentation and classification model, the training set consisted of 100 patients and evaluated on a testing set of 50 patients. The segmentation model was trained by us for 300 epochs along with a batch size of 4 as well as an initial learning rate of $5e-4$ for segmentation training. In order to enhance convergence, a learning rate scheduler that employs cosine annealing was implemented. An NVIDIA A100 GPU was employed to perform segmentation and classification. The complete pipeline, which included pre-processing, segmentation, post-processing, feature extraction, and two-stage classification, required an estimated 3 seconds per patient, indicating a significant potential for integration into our platform clinical workflows. Figure~\ref{fig:demo_output} shows the ground truth and corresponding segmented images and classification result by our model.

\textbf{Segmentation Results:} Table~\ref{tab:Results_segmentation} compares the segmentation performance of our 3D residual U-Net model at variance with recent state-of-the-art methods. Our model achieved an average DSC of \textbf{92.56\%}, outperforming several recent state-of-the-art methods including TransUNet~\cite{Chen2021TransUNet}, Swin-Unet~\cite{Cao2021SwinUnet}, SegFormer3D~\cite{Perera2024}, and even the widely used nnUNet~\cite{Isensee2021nnUNet} and nnFormer~\cite{Zhou2021nnFormer} baselines. Specifically, IntelliCardiac achieved the highest segmentation accuracy for the RV (92.27\%) and Myo (90.33\%), and maintained competitive performance for the LV (95.09\%). These results show that our method is highly effective in capturing both global and local anatomical structures, particularly in challenging regions like the RV and myocardial boundaries. The strong segmentation performance provides a reliable foundation for downstream feature extraction and disease classification within the IntelliCardiac pipeline.

IntelliCardiac's exceptional result is because of numerous innovations. We present a logical-depth cropping strategy that is tailored to the temporal depth of the image volume. Our method dynamically accounts for variations in cardiac images slice depth by generating multiple crops when the temporal dimension exceeds the target size, in contrast to conventional cropping methods that extract a single fixed-volume region. Apart from improving resilience to different scan lengths, this helps the model to leverage several spatial-temporal viewpoints. When the depth is small, symmetric padding ensures consistency free from information loss. This adaptive approach assures complete coverage of the cardiac cycle and enhances generalization over patients using different acquisition techniques. Another contribution is our dynamically weighted Focal Dice Loss, which modifies weights in real-time according to past epoch performance. This immediately address class imbalance and improve segmentation uniformity, this encourages the model to give priority to minority or underperforming buildings. The anatomical coherence is further refined by post-processing with Largest Connected Component Analysis (LCCA) to eliminate isolated false positives. Collectively, these methodological advancements yield consistent improvements in all assessed metrics, thereby establishing a dependable and precise foundation for downstream disease classification within the IntelliCardiac pipeline.

Although IntelliCardiac demonstrated slightly lower Dice score for the left ventricle (95.09\%) compared to some segmentation-focused baselines may reflect a purposeful redistribution of model attention toward more diagnostically critical regions. The myocardium, in particular, showed a substantial improvement (90.33\%), which is especially important given its thin geometry and ambiguous boundaries. Because myocardial thickness, shape, and motion patterns are essential for downstream disease classification, especially in distinguishing myocardial infarction (MINF) from dilated cardiomyopathy (DCM), this improvement is clinically meaningful. Altogether, these architectural and algorithmic innovations yield consistent gains across all metrics while strategically prioritizing accuracy in regions that most directly support diagnostic decision-making within the IntelliCardiac pipeline.

\textbf{Classification Results:} Beyond segmentation, we evaluated the performance of our disease diagnosis module, which classifies patients into one of five cardiac disease categories. As shown in Table~\ref{tab:Results_seg_clas}, IntelliCardiac outperformed several existing methods in both segmentation and classification tasks. Specifically, it achieved a classification accuracy of \textbf{98\%}, exceeding the performance of previous works by Wolterink~\cite{Wolterink2018}, Wibowo~\cite{Wibowo2022}, Zheng~\cite{Zheng2019}, and Khened~\cite{Khened2018}, as well as more recent approaches that incorporate handcrafted features, motion-based modeling, or dense segmentation backbones. This improvement is attributed to our two-stage classification strategy, which first employs a Random Forest classifier trained on global anatomical and functional features, and then applies an expert-level SVM classifier to refine predictions between closely related conditions like MINF and DCM. 

Figure~\ref{fig:confusion_matrices} visualizes the classification outcomes before and after expert refinement. The initial classifier model obtained 92\% accuracy; however, it displayed confusion between MINF and DCM. The final accuracy was 98\% after the application of advanced classifier correction using two features based on myocardial wall thickness: the mean of slice-wise standard deviations of MWT and the standard deviation of slice-wise mean MWT at end-diastole. These two orthogonal but related characteristics are intended to capture inter-regional variability and segment-level heterogeneity, which are important indicators for differentiating between diffuse thinning in DCM and focal thinning in MINF. By focusing only on these two characteristics, IntelliCardiac avoids minimizes overfitting and attains robust generalization. When the feature used in conjunction with an RBF-kernel SVM and limited to ambiguous cases, our classification module can surpass other models.
Together, these results demonstrate the effectiveness of IntelliCardiac as a fully integrated platform capable of delivering high-accuracy segmentation and robust cardiac disease classification from cine MRI scans.

\subsection{Ablation study}
We performed an ablation study using a variety of segmentation backbone configurations to assess the effect of various loss functions on model performance. The results of the study are presented in Table~\ref{tab:Results_ablation_study}, which includes the classification accuracy for disease prediction, as well as the average Dice score (DSC) for three anatomical structures: the myocardium, the right ventricle, and the left ventricle. We compared standard Cross Entropy~\cite{ronneberger2015u}, Dice~\cite{Sudre2017Dice}, and Focal Dice Loss~\cite{Wang2018} against our proposed dynamically weighted Focal Dice Loss. The Cross Entropy loss had an average DSC of 92.21\% and a classification accuracy of 96\%. Substituting with Dice loss improved segmentation (92.45\%). The Focal Dice loss yielded the same average Dice score but resulted in a drop in classification performance.

Our proposed loss function achieved the highest segmentation performance (92.56\%), surpassing all baselines, and led to the best classification result with a final accuracy of 98\%. To assess the impact of anatomical localization, we also evaluated our model without Region of interest (ROI) detection. The absence of ROI cropping led to a performance drop (92.17\%), and final classification accuracy declined to 94\%. These results demonstrate that both the proposed loss function and ROI-based preprocessing contribute significantly to segmentation consistency and improved downstream disease classification.
\section{Conclusion}
In this work, we designed \textbf{IntelliCardiac}, a web-based application that provides real-time visualization and AI-assisted cardiac diagnostic help with an easy interface for patients, cardiologists, and physicians. In this application study, we achieve a mean Dice score of 92.6\% through cardiac structural segmentation. Then extract features from segmentation outputs to achieve 98\% precision in disease classification across five diagnostic categories. These results consistently outperform state-of-the-art segmentation diagnosis methods. IntelliCardiac demonstrates that deep learning and domain expertise can provide robust, accurate, and scalable cardiac imaging tools. We aim to extend the platform’s generalizability to other medical image datasets and pathologies, and explore deployment across diverse clinical environments. IntelliCardiac's high performance and user-centered design could improve automated, interpretable, and accessible cardiac treatment.
\section*{Acknowledgment}
The authors appreciate the computational resource provided by the University at Albany -- SUNY. 

\bibliographystyle{plain}

\bibliography{reference}

\end{document}